\documentclass[11pt, 3p, authoryear]{elsarticle}
\pdfoutput=1
\usepackage{natbib,graphics,amsmath,amsfonts,amssymb,tabularx,geometry,multirow,diagbox,xcolor, longtable,enumerate,amsthm,hyperref}
\usepackage{hyphenat}
\usepackage[english]{babel}
\newcommand{\pcite}[1]{\citeauthor{#1}'s (\citeyear{#1})}

\journal{arXiv}

\title{On Long Memory Origins and Forecast Horizons}
\author{J. Eduardo Vera-Vald\'es\corref{fn1}}
\address{Department of Mathematical Sciences, Aalborg University, and CREATES.}
\cortext[fn1]{E-mail: \href{mailto:eduardo@cimat.mx}{eduardo@math.aau.dk} Webpage: \url{https://sites.google.com/view/veravaldes/} \\ Address: Skjernvej 4A, 9220 Aalborg East, Denmark. } 

\date{\today}

\newcommand{\tabitem}{~~\llap{\textbullet}~~}
\allowdisplaybreaks

\begin{document}
	
	\begin{abstract}
		\noindent Most long memory forecasting studies assume that the memory is generated by the fractional difference operator. We argue that the most cited theoretical arguments for the presence of long memory do not imply the fractional difference operator, and assess the performance of the autoregressive fractionally integrated moving average $(ARFIMA)$ model when forecasting series with long memory generated by nonfractional processes. We find that high-order autoregressive $(AR)$ models produce similar or superior forecast performance than $ARFIMA$ models at short horizons. Nonetheless, as the forecast horizon increases, the $ARFIMA$ models tend to dominate in forecast performance. Hence, $ARFIMA$ models are well suited for forecasts of long memory processes regardless of the long memory generating mechanism, particularly for medium and long forecast horizons. Additionally, we analyse the forecasting performance of the heterogeneous autoregressive ($HAR$) model which imposes restrictions on high-order $AR$ models. We find that the structure imposed by the $HAR$ model produces better long horizon forecasts than $AR$ models of the same order, at the price of inferior short horizon forecasts in some cases. Our results have implications for, among others, Climate Econometrics and Financial Econometrics models dealing with long memory series at different forecast horizons. We show in an example that while a short memory autoregressive moving average $(ARMA)$ model gives the best performance when forecasting the Realized Variance of the S\&P 500 up to a month ahead, the $ARFIMA$ model gives the best performance for longer forecast horizons.
		
		\bigskip
		
		
		
		
	\end{abstract}
	
\maketitle

\clearpage

\section{Introduction}

Long memory analysis deals with the notion of series with strong persistence in the sense of long lasting correlations. One of the first works on strong persistence is due to \cite{Hurst1956}. He studied the long-term capacity of reservoirs for the Nile and recommended to increase the height of a dam to be built given his observations on cycles of highs at the river. As found by Hurst, failing to account for the presence of long memory can lead to inaccurate forecasts. If the data is best modelled by a long memory process, then the predictions computed with standard models would be too optimistic, in the sense that they would predict a return to normal events faster than what we would observe in reality. A dam built based on a short memory forecast would be more prone to overflow that one built based on a long memory forecast, hence increasing the risk of a catastrophic event. Hurst's work highlights the importance of developing appropriate forecasting tools to deal with the presence of long memory.

In the time series literature, the $ARFIMA$ class of models remains to be the most popular given its appeal of bridging the gap between the stationary $ARMA$ models, and the nonstationary $ARIMA$ model. Moreover, some effort has been directed to assess the performance of the $ARFIMA$ type of models when forecasting long memory processes. 

\cite{Ray1993} calculates the percentage increase in mean-squared error ($MSE$) from forecasting fractionally integrated $(FI)$ series with $AR$ models. She argues that the $MSE$ may not increase significantly, particularly when we do not know the true long memory parameter. \cite{Crato1996} compare the forecasting performance of $ARFIMA$ models against $ARMA$ alternatives and find that $ARFIMA$ models are in general outperformed by $ARMA$ alternatives for short forecast horizons. Looking at real data, \cite{Martens2009} show that for daily realized volatility for forecast horizons of up to twenty days, it seems to be beneficial to use a flexible high-order $AR$ model instead of a parsimonious but stringent fractionally integrated model. On the other hand, \cite{Barkoulas1997} find improvements in forecasting accuracy when fitting $ARFIMA$ models to Eurocurrency returns series, particularly for longer horizons. By allowing for larger data sets of both financial and macro variables, and considering larger forecast horizons, \cite{Bhardwaj2006} find that $ARFIMA$ processes generally outperform $ARMA$ alternatives in terms of forecasting performance. Thus, there does not seem to be a consensus regarding the forecast performance of the $ARFIMA$ model.

One thing that most forecasting comparison studies have in common is the underlying assumption that long memory is generated by an $ARFIMA$ process. There are two predominant theoretical explanations for the presence of long memory in the time series literature: cross-sectional aggregation of dynamic persistent micro units (\citealp{Granger1980}), and that shocks may be of random duration (\citealp{Parke1999}). As argued in Section \ref{Sec:LMGP}, neither of these sources of long memory imply an $ARFIMA$ specification. The question addressed in this paper is if an $ARFIMA$ specification serves as a good approximation for forecasting purposes when the long memory generating mechanism is different from the $ARFIMA$ model. 

Moreover, as argued by \cite{Baillie2012}, a practitioner's goals will generally include making forecasts over both short and long horizons. As an example, the surge of Climate Econometrics as a way to address Climate Change relies on the construction of long horizon forecasts while addressing medium term policy goals. Thus, we analyse the forecasting performance of short and long memory models at several forecast horizons. In particular, we extend previous studies to larger forecast horizons relevant to Climate Change analysis. 

This paper proceeds as follows. In Section \ref{Sec:LMGP}, we present the long memory generating processes considered, and show that the most cited theoretical explanations for the presence of long memory do not imply an $ARFIMA$ specification. Section \ref{Sec:MC} describes the design of the Monte Carlo analysis used for the forecasting study. Section \ref{Sec:Res} presents the results from the forecasting analysis, while Section \ref{Sec:Disc} discusses them in a bias-variance trade-off context. Moreover, Section \ref{Sec:App} shows that the insights gained from the Monte Carlo simulations hold on real data. Finally, Section \ref{Sec:Con} presents the conclusions.

\section{Long Memory Generating Processes}\label{Sec:LMGP}

In this section, we present the selected processes used to generate long memory. All processes considered are long memory in the covariance sense, see \cite{Haldrup2017} for other definitions. In contrast to the alternatives, the covariance sense relates to the rate of decay of the autocorrelations. In this sense, the fitted models try to mimic the rate of decay of the weight that past observations have on future realizations. In this context, the models this information to produce better forecasts; thus, the covariance sense is a sensible definition of long memory for forecasting purposes.

\subsection{The {\textit{ARFIMA}} Model}

As a benchmark, we include the $ARFIMA$ process due to \cite{Granger1980b}, and \cite{Hosking1981} in the analysis. The authors extended the $ARMA$ model to include fractional dynamics by considering the process
\begin{equation}\label{eq:arfima}
\phi(L)(1-L)^d x_t = \theta(L)\epsilon_t,
\end{equation}
where $\epsilon_t$ is a white noise process, $d\in(-1/2,1/2)$, and $\phi(L)$ and $\theta(L)$ are polynomials in the lag operator with no common roots, all outside the unit circle. The authors used the standard binomial expansion to decompose the fractional difference operator $(1-L)^d$ in a series with coefficients $\pi_j=\Gamma(j+d)/(\Gamma(d)\Gamma(j+1))$ for $j\in\mathbb{N}$. Using Stirling's approximation, it can be shown that these coefficients decay at a hyperbolic rate, which in turn translates to slowly decaying autocorrelations.

It is well known that $ARFIMA$ processes are long memory by all definitions typically considered in the literature, and are relatively easy to estimate by Maximum Likelihood, see \cite{Sowell1992}. Thus, the $ARFIMA$ model has become the canonical construction for modelling and forecasting long memory in the time series literature; see \cite{Beran1994}, and \cite{Baillie1996} for a review. 

For the Monte Carlo analysis, we consider $ARFIMA(1,d,0)$ processes as a way to incorporate both long and short term dynamics. 

\subsection{Cross-Sectional Aggregation}

\cite{Granger1980}, in line with the work of \cite{Robinson1978} on autoregressive processes with random coefficients, showed that aggregating $AR(1)$ processes with coefficients sampled from a Beta distribution can produce long memory. He considered $N$ series generated as
$$x_{i,t} = \alpha_i x_{i,t-1}+\varepsilon_{i,t}\ \ \ i=1, 2, \cdots, N;$$
where $\varepsilon_{i,t}$ is a white noise process with $E[\varepsilon_{i,t}^2] = \sigma_\varepsilon^2$ $\forall i \in\{1,2,\cdots,N\}$, $\forall t \in \mathbb{Z}$. Moreover, $\alpha_i^2 \sim \mathcal{B}(\alpha; p, q)$ with $p,q>1$, and where $\mathcal{B}(\alpha; p, q)$ is the Beta distribution with density given by
$$\mathcal{B}(\alpha; p, q) =  \frac{1}{B(p,q)} \alpha^{p-1}(1-\alpha)^{q-1}\ \ \ \ \text{for}\ \ \ \alpha\in(0,1),$$
with $B(\cdot,\cdot)$ the Beta function. Furthermore, define the cross-sectional aggregated series as
$$x_t = \frac{1}{\sqrt{N}}\sum_{i=1}^N x_{i,t}.$$

Granger showed that as $N\to\infty$, the autocorrelations of $x_t$ decay at a hyperbolic rate with parameter $d=1-q/2$; thus, $x_t$ has long memory in the covariance sense.

The cross-sectional aggregation result has been extended in different ways, including to allow for general $ARMA$ processes, and to other distributions; see \cite{Oppenheim2004}, \cite{Linden1999}, and \cite{Zaffaroni2004}.

\cite{Haldrup2017} showed that the long memory generated by cross-sectional aggregation does not correspond to the one associated to the $ARFIMA$ model. In particular, they showed that although the long memory by cross-sectional aggregation can be removed by fractional differencing, the resulting series does not belong to the class of linear $ARMA$ processes. The question addressed in this paper is whether an $ARFIMA$ specification remains useful for forecasting purposes.


\subsection{Error Duration Model}

The error duration model was introduced by \cite{Parke1999}. He showed that if the series is the result of the sum of shocks of stochastic duration, then it would exhibit long memory in the form of hyperbolic decaying autocorrelations. 

Let $\varepsilon_s$ be a series of $i.i.d.$ shocks with mean zero and finite variance $\sigma^2$. Assume that the shock $\varepsilon_i$ has a stochastic duration of $n_i\geq0$ time periods, and thus surviving from period $i$ until period $i+n_i$. Let $p_k$ be the probability that event $\varepsilon_i$ survives until period $i+k$, and take $g_{i,t}$ to be the indicator function for the event that the error $\varepsilon_i$ survives until period $t$. Furthermore, define $x_t$ as
$$x_t = \sum_{s=-\infty}^t g_{s,t}\varepsilon_s.$$
Then, if\footnote{For two series $a_t,b_t$, with $b_t\neq 0$ $\forall t$, we write $a_t\sim b_t$ if $\lim_{t\to\infty}a_t/b_t = 1$.} $p_k\sim k^{-2+2d}$ as $k\to\infty$, $x_t$ will have long memory in the covariance sense.

By properly choosing the error survival probabilities, Parke showed that the autocorrelation function will decay at a rate similar to $FI(d)$ processes. However, the resulting series has dichotomous coefficients that do not correspond to the fractional difference operator. 

We follow Parke's specification in the Monte Carlo analysis and consider error survival probabilities that mimic those of the $FI(d)$ model. 

Table \ref{tab:lmgp} summarizes the long memory generating mechanisms to be analysed.

\begin{small}
	\begin{longtable}{l|c}\caption{Long Memory Generating Processes}\label{tab:lmgp}\\
		\hline
		\hline
		\begin{tabular}{l} $ARFIMA(p,d,q)$ \\ ($DGP$ 1) \end{tabular} &$\begin{aligned} \phi(L)(1-L)^d x_t &= \theta(L)\varepsilon_t \\ \phi(z)&=1-\phi_1z-\cdots-\phi_pz^p \\
		\theta(z) &= 1+\theta_1z+\cdots+\theta_qz^q \\ (1-L)^d &=\sum_{s=0}^\infty{ \frac{\Gamma(s-d)}{\Gamma(-d)\Gamma(s+1)}L^s} \\ \end{aligned}$\\
		\hline
		\begin{tabular}{l} Cross-Sectional Aggregation \\ ($DGP$ 2) \end{tabular} &$\begin{aligned} x_t &= \frac{1}{\sqrt{N}}\sum_{i=1}^N x_{i,t}\\ x_{i,t} &= \alpha_i x_{i,t-1}+\varepsilon_{i,t}\\ \alpha_i&\sim \mathcal{B}(\alpha; p, q);\ p,q>1\\ \end{aligned}$ \\
		\hline
		\begin{tabular}{l} Error Duration Model \\ ($DGP$ 3) \end{tabular}	&$\begin{aligned} x_t &= \sum_{s=-\infty}^t g_{s,t}\varepsilon_s \\ g_{s,s+k} &= \left\{\begin{tabular}{ll} 0 &w.p. $1-p_k$\\ 1 &w.p. $p_k$\end{tabular} \right. \\ p_k &= k^{2d-2}\\ \end{aligned}$ \\
		\hline
	\end{longtable}
\end{small}

\section{Monte Carlo Design}\label{Sec:MC}

In this section, we describe the Monte Carlo analysis designed to compare the forecasting performance of $ARFIMA$ models against $ARMA$ and high-order $AR$ models on long memory series generated by the processes described in Section \ref{Sec:LMGP}. 

\subsection{Forecast Evaluation}\label{Subsec:MCS}
We use the Model Confidence Set ($MCS$) approach of \cite{Hansen2011} to assess the forecasting performance of the selected models. From an initial set of models, the methodology allows us to obtain the superior set at a given confidence level. In this sense, the $MCS$ is better suited to compare the forecast performance of a large set of competing models.

The $MCS$ algorithm proceeds as follows. From a starting set of competing model, $\mathcal{M}_0$, we search for the set of superior models at forecast horizon $h$, $\mathcal{M}^*$, defined by
$$\mathcal{M}^*=\{i\in\mathcal{M}_0\ |\ E(d_{i,j}^h)\leq0\ \ \ \forall j\in \mathcal{M}_0\},$$
where $d_{i,j}^h$ is the loss differential between models $i$ and $j$. 

We obtain $\mathcal{M}^*$ by sequential elimination. For each long memory generating process, we fit all the competing models in the starting set for a sample size $T$. The models are indexed by $i\in\{1,2,\ldots,m\}$, and the out of sample forecast from model $i$ is denoted by $\hat{y}^i_{T+k}$, $\forall k\in\{1,\ldots,h\}$. We rank the models according to their expected loss using one of two loss functions: the mean square error ($MSE$), $L_{SQ}\left(y_{T+k},\hat{y}^i_{T+k}\right) = \left(y_{T+k}-\hat{y}^i_{T+k}\right)^2$, and the mean absolute deviation ($MAD$), $L_{AD}\left(y_{T+k},\hat{y}^i_{T+k}\right) = \left|y_{T+k}-\hat{y}^i_{T+k}\right|$. 

Define the loss differential between models $i$ and $j$ by
$$d_{i,j}^k = L_{M}\left(y_{T+k},\hat{y}^i_{T+k}\right) - L_{M}\left(y_{T+k},\hat{y}^j_{T+k}\right),$$
for $M=SQ,AD$; $i,j\in\{1,2,\ldots,m\}$. We eliminate the worst performing model at each step, and we continue with the process until we can not reject the null hypothesis of equal loss differentials for all models in the set; that is, 
$$H_0:E(d_{i,j}^k)\leq0\ \ \ \forall i,j\in \mathcal{M}.$$

The null is tested by using either the range statistic, $T_R$, or the semiquadratic statistic, $T_{SQ}$, defined by $$T_R = \max_{i,j\in\mathcal{M}}\frac{|\bar{d}_{i,j}|}{\left(\widehat{var}(\bar{d}_{i,j})\right)^{1/2}}\ \ \ \ \ \ T_{SQ} = \sum_{i\neq j}{\frac{(\bar{d}_{i,j})^2}{\left(\widehat{var}(\bar{d}_{i,j})\right)^{1/2}}}.$$

In the Monte Carlo analysis, we present the percentage number of times each model is contained in $\mathcal{M}^*$ for each forecast horizon.

Additionally, as another measure of forecast performance, we compute both the out of sample root mean square error ($RMSE$), and the out of sample root mean absolute deviation ($RMAD$) given by
$$RMSE_h^i = \left(\frac{1}{h}\sum_{k=1}^h{\left(y_{T+k}-\hat{y}^i_{T+k}\right)^2}\right)^{1/2}\ \ \ \ RMAD_h^i = \left(\frac{1}{h}\sum_{k=1}^h{\left|y_{T+k}-\hat{y}^i_{T+k}\right|}\right)^{1/2},$$
where $h$ and $\hat{y}^i_s$ are defined as above. We report the mean of both $RMSE$ and $RMAD$ across all replications.

Note that the $MCS$, and $RMSE$ or $RMAD$ evaluation criteria are complementary by construction. The $MCS$ measure computed in this way, the percentage number of times each model is contained in the set of superior models, tells us about the success rate of the models; that is, how often do we expect each model to perform well. Meanwhile, the $RMSE$ and $RMAD$ criteria measure the average performance of each model. We will see in Section \ref{Sec:Res} how this distinction becomes relevant when selecting a forecasting model. 

\subsection{Model Selection}

This section presents the models considered for the forecasting analysis. Table \ref{tab:competitors} presents the starting set, $\mathcal{M}_0$, for the $MCS$ approach explained in Section \ref{Subsec:MCS}.

\begin{table}[h!]
	\centering
	\begin{small}
		\caption{Starting Set $\mathcal{M}_0$}
		\begin{tabular}{ccc}
			\hline
			\hline
			$FI(d)$	&$ARMA(1,1)$ &$HAR(3)$\\
			$ARFIMA(1,d,0)$	&$ARMA(2,1)$ &$AR(22)$\\
			$ARFIMA(0,d,1)$	&$ARMA(1,2)$ &$AR(30)$\\
			$ARFIMA(1,d,1)$	&$ARMA(3,3)$ &$AR(50)$\\
			$ARFIMA(2,d,1)$	&$ARMA(4,4)$ &$I(1)$\\
			\hline
		\end{tabular}\label{tab:competitors}
	\end{small}
\end{table}

Model selection was based on two criteria.

As a first criterion, we use the Bayesian Information Criterion ($BIC$) to select the number of lags to include in both the $ARFIMA$ and $ARMA$ models in an independent Monte Carlo analysis. The validity of the $BIC$ for the class of processes with fractional differencing was proven by \cite{Beran1998}. The authors show that for this class of processes the penalty term must tend to infinity simultaneously with the sample size; thus, the Akaike Information Criterion is not consistent while the $BIC$ is. Note that we made the lag selection exercise independent from the forecasting analysis to avoid the multiple testing problem.

We allow for a maximum of two lags at both components of the $ARFIMA$ model, while the maximum was set to four for the $ARMA$ model. We use Maximum Likelihood for the estimation of both classes of models with parameter specifications as reported in Appendix \ref{app:par}.

Results from the lag selection exercise, presented in Appendix \ref{app:lags}, show that not many lags are selected for the $ARFIMA$ specification for either component. This suggests that the short term component is not that persistent once we control for the long memory behaviour. For the $ARMA$ specification, perhaps not surprisingly, more lags are selected due to the fact that we are not controlling for the long memory behaviour by way of the estimation of the fractional memory $d$. Nonetheless, the maximum number of lags selected by the $BIC$ is two.

As a second criterion, in addition to the preferred models from the lag selection exercise, we follow previous works on long memory forecasting and consider high-order $AR$ processes, $AR(30)$ and $AR(50)$. Moreover, given the success of the $HAR(3)$ model of \cite{Corsi2009} on mimicking long memory behaviour, see for instance \cite{Andersen2007} and \cite{Chiriac2011}, we include both the unconstrained $AR(22)$, and the $HAR(3)$ models. The $HAR(3)$ model is a constrained $AR(22)$ given by $$x_t = a_0+a_1x_{t-1}^{(f)}+a_2 x_{t-1}^{(w)} + a_3 x_{t-1}^{(m)}+\epsilon_t,$$ where $x_{t-1}^{(f)}=x_{t-1}$, $x_{t-1}^{(w)}=\frac{1}{5}\sum_{i=1}^5{x_{t-i}}$ and, $x_{t-1}^{(m)}=\frac{1}{22}\sum_{i=1}^{22}{x_{t-i}}$. 

The $HAR$ specification has been used to model financial data, it reflects the fact that different agents respond to uncertainty at distinct horizons. In this context, the three components of the model seek to capture the daily $(x_t^{(f)})$, weekly $(x_t^{(w)})$, and monthly $(x_t^{(m)})$ levels of uncertainty. 

Note that including the $HAR(3)$ model allows us to extend \pcite{Corsi2009} results in several directions. We make comparisons against a larger set of models, and we use the $MCS$ approach, which is better suited for comparisons between multiple alternatives. Also, we include larger forecast horizons, and we remove the uncertainty regarding the presence of long memory in the data by comparing the performance of the $HAR$ model in simulated long memory series, whereas Corsi used real data.

\subsection{Monte Carlo Design}

All models were estimated by Maximum Likelihood ($MLE$) following the work of \cite{Baillie2012} on long memory estimators for forecasting purposes. The authors find the forecasts based on $MLE$ to be superior than the ones obtained from local Whittle estimators. Moreover, throughout, we use a large sample size of $T=1,000$ to reduce the estimation error, and we consider values of the long memory parameter in the stationary range, $d\in(0,1/2)$. Furthermore, given the rise of Climate Econometrics studies keen on producing long horizon forecasts, we consider it relevant to evaluate forecast performances to horizons as far as $h=300$, which correspond to twenty-five years of monthly forecasts.

Table \ref{tab:MC_design} presents the Monte Carlo design for the forecasting analysis.

\begin{table}[ht!]
	\centering
	\caption{Monte Carlo Design}
	\begin{tabular}{l}
		\hline
		\hline
		\tabitem Generate series of size $T+h$ using the long memory generating processes considered,\\ 
		Section \ref{Sec:LMGP}, Table \ref{tab:lmgp}. The model calibrations are reported in Table \ref{tab:params}
		in Appendix \ref{app:par}.\\
		\tabitem Fit by Maximum Likelihood the competing models in the starting set $\mathcal{M}_0$,
		Table \ref{tab:competitors},\\for a sample size $T$.\\
		\tabitem Construct forecasts from each model for horizons $h\in\{5,10,30,50,100,300\}$.\\
		\tabitem Determine the $MCS$ and compute the $RMSE$ and $RMAD$.\\
		\tabitem Repeat the steps above $R$ times, the number of replications.\\
		\hline
		\tabitem After the $R$ replications, report the percentage number of times each model is contained in\\the $MCS$, and the mean values of $RMSE$ and $RMAD$, for each forecast horizon.\\
		\hline
	\end{tabular}\label{tab:MC_design}
\end{table}

\vspace*{2cm}
\section{Monte Carlo Results}\label{Sec:Res}

In this section, we present the results from the Monte Carlo simulations. The parameters for the simulations are presented in Appendix \ref{app:par}. Throughout, for reasons of space, we focus on the $MAD$ loss function given that it is less sensitive to large misspredictions, see \cite{Hansen2003}. Nonetheless, tables using the $MSE$ loss function, reported in an Online Appendix, show similar results. 

\subsection{{\textit{DGP}} 1: ARFIMA}

\begin{figure}[h!]\caption{Proportion of times the top performing models are in the $MCS$ at a 95\% confidence level when forecasting $DGP$ 1 with different degrees of memory at several horizons.}\label{plot:DGP1}
	\hspace*{-1cm}   
	\includegraphics[scale=0.45]{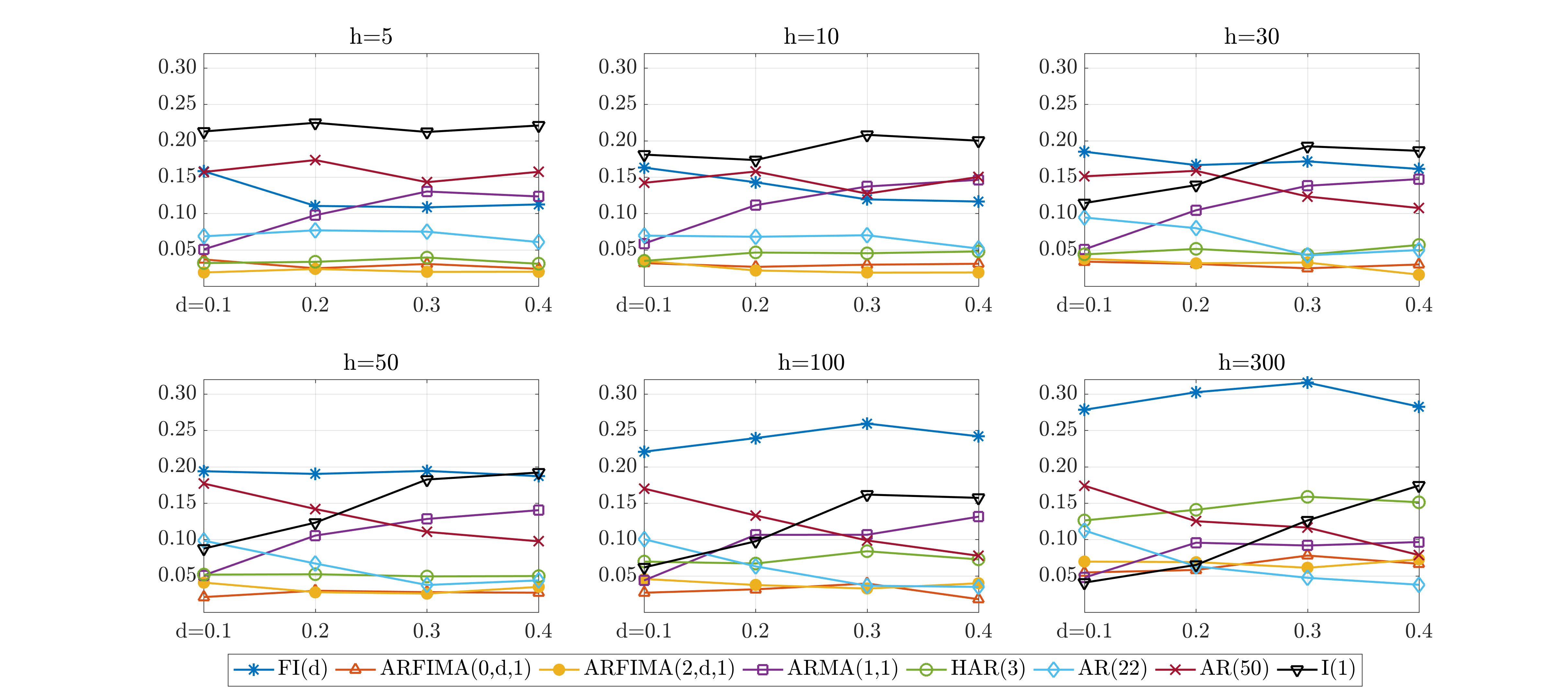}
	\vspace*{-0.4cm}
\end{figure}

As a benchmark, we present in Table \ref{tab:DGP1} and Figure \ref{plot:DGP1} the results from the Monte Carlo analysis for an $ARFIMA(1,d,0)$ process, $DGP$ 1, for $d=0.3$.

\begin{table}[h!]
	\centering
	\begin{scriptsize}
		\setlength{\tabcolsep}{3pt}
		\renewcommand{\arraystretch}{1} 
		\begin{tabular}{l|cc|cc|cc|cc|cc|cc}
			\hline
			$DGP$ 1	&\multicolumn{2}{c|}{h=5}	&\multicolumn{2}{c|}{10}		&\multicolumn{2}{c|}{30}		&\multicolumn{2}{c|}{50}	&\multicolumn{2}{c|}{100}	&\multicolumn{2}{c}{300}	\\
			$d=0.3$	&	$RMAD$	&	$MCS$	&	$RMAD$	&	$MCS$	&	$RMAD$	&	$MCS$	&	$RMAD$	&	$MCS$ &	$RMAD$	&	$MCS$ &	$RMAD$	&	$MCS$	\\
			\hline																	
			$FI(d)$		&	0.937	&	0.109	&	0.964	&	0.118	&	0.984	&	0.171	&	0.988	&\textbf{0.193}	&	0.995	&\textbf{0.259}	&	1.004	&\textbf{0.315}	\\
			$ARFIMA(1,d,0)$&	\textbf{0.933}	&	0.029	&	0.962	&	0.028	&	0.983	&	0.026	&	0.988	&	0.034	&\textbf{0.994}	&	0.035	&\textbf{1.003}	&	0.055	\\
			$ARFIMA(0,d,1)$&	\textbf{0.933}	&	0.031	&\textbf{0.961}	&	0.030	&\textbf{0.982}	&	0.025	&\textbf{0.987}	&	0.028	&\textbf{0.994}	&	0.039	&\textbf{1.003}	&	0.078	\\
			$ARFIMA(1,d,1)$&	0.935	&	0.009	&	0.963	&	0.011	&	0.983	&	0.018	&	0.988	&	0.028	&	\textbf{0.994}	&	0.025	&\textbf{1.003}	&	0.053	\\
			$ARFIMA(2,d,1)$&	0.935	&	0.020	&	0.963	&	0.019	&	0.984	&	0.033	&	0.989	&	0.026	&	0.995	&	0.033	&	1.004	&	0.061	\\
			\hline
			$ARMA(1,1)$	&	0.944	&	0.130	&	0.975	&	0.137	&	0.995	&	0.138	&	0.999	&	0.128	&	1.002	&	0.107	&	1.007	&	0.092	\\
			$ARMA(2,1)$	&	0.937	&	0.023	&	0.966	&	0.027	&	0.988	&	0.013	&	0.993	&	0.025	&	0.999	&	0.024	&	1.006	&	0.045	\\
			$ARMA(1,2)$	&	0.938	&	0.036	&	0.968	&	0.032	&	0.990	&	0.039	&	0.995	&	0.037	&	1.000	&	0.034	&	1.006	&	0.043	\\
			$ARMA(3,3)$	&	0.937	&	0.032	&	0.967	&	0.034	&	0.988	&	0.026	&	0.993	&	0.019	&	0.999	&	0.025	&	1.007	&	0.048	\\
			$ARMA(4,4)$	&	0.938	&	0.040	&	0.967	&	0.042	&	0.988	&	0.035	&	0.993	&	0.037	&	0.999	&	0.030	&	1.007	&	0.063	\\
			\hline
			$HAR(3)$		&	0.936	&	0.039	&	0.965	&	0.045	&	0.986	&	0.043	&	0.992	&	0.049	&	0.998	&	0.084	&	1.006	&	0.159	\\
			$AR(22)$		&	0.939	&	0.075	&	0.967	&	0.070	&	0.988	&	0.042	&	0.992	&	0.038	&	0.998	&	0.037	&	1.005	&	0.047	\\
			$AR(30)$		&	0.941	&	0.073	&	0.970	&	0.072	&	0.989	&	0.076	&	0.993	&	0.067	&	0.998	&	0.060	&	1.005	&	0.071	\\
			$AR(50)$		&	0.948	&	0.143	&	0.976	&	0.127	&	0.994	&	0.123	&	0.997	&	0.111	&	1.000	&	0.099	&	1.006	&	0.116	\\
			$I(1)$		&	1.036	&\textbf{0.212}	&	1.076	&\textbf{0.208}	&	1.113	&\textbf{0.193}	&	1.125	&	0.183	&	1.151	&	0.162	&	1.200	&	0.126	\\
			\hline																
		\end{tabular}
		\caption{Mean of the $RMAD$ and proportion of times the model is in the $MCS$ using the $MAD$ loss function and the $T_R$ statistic at a 95\% confidence level.}
		\label{tab:DGP1}
	\end{scriptsize}
\end{table}

Table \ref{tab:DGP1} shows that $ARFIMA$ models are the preferred specification for all forecast horizons measured by the $RMAD$ criterion, which is not surprising given that $DGP$ 1 is indeed an $ARFIMA$ process. Turning to the $MCS$ criterion, note that the no-change $I(1)$ model gives the best forecast performance for short horizons, while the $FI(d)$ model is the preferred one for medium and large forecast horizons, and its relative performance increases with the forecast horizon. 

The results for the $I(1)$ model are of particular interest. Note that it is the preferred model by the $MCS$ criterion for short forecast horizons, while it gives the worst performance by the $RMAD$ criterion for all horizons. As discussed in Section \ref{Subsec:MCS}, these apparent conflicting results can be explained given the complimentary nature of the forecast evaluation measures. Recalling that the $RMAD$ is a measure across all replications, the results suggest that when the $I(1)$ is not in the model confidence set, its forecasts perform badly. Nonetheless, its success rate for short forecast horizons could make it a reasonable alternative in some cases, which will probably relate to the distance from the last observation before the forecast to the overall mean. If the last observation is already near the mean, and given that all models will forecast a return to the mean given the stationarity assumption, the no-change model could provide a good forecast alternative, the mean.

Looking at both criteria, we find that high-order $AR$ and $ARMA$ models perform quite well when forecasting a true $ARFIMA$ process for short forecast horizons. In particular, the $AR(50)$ and $ARMA(1,1)$ fall in the superior set of models more than times than $ARFIMA$ specifications for $h=5$, and $h=10$; while the increase in the $RMAD$ criterion relative to $ARFIMA$ specifications is not too large. A practitioner could in principle construct a weighted measure between the two criteria depending on the problem at hand and choose to use this class of models.

Nonetheless, for medium and large forecast horizons, the $FI(d)$ model is the one contained in the $MCS$ the most, and with a $RMAD$ close to the minimum. In this sense, the $FI(d)$ appears to be a good overall model for forecasts at medium and large horizons. The superior performance of the $FI(d)$ model compared to the correct $ARFIMA(1,d,0)$ specification may be explained given the small value of the autoregressive coefficient and the estimation error in the long memory parameter. The table suggests that the $FI(d)$ model seems to capture enough information for forecasting purposes in the long horizon once the short memory component fades away.\footnote{Results allowing more short-term dynamics are presented in the Online Appendix.} 

Furthermore, Figure \ref{plot:DGP1} allows us to contrast the performance of high-order $AR$ models and $ARFIMA$ models to different degrees of memory.\footnote{For ease of exposition, we present a subset of the top performing models in the figures; nonetheless, we present plots with all competing models in the Online Appendix.} The figure shows that for $h=5$ and $h=10$, and for all degrees of memory, the $AR(50)$ produces better or similar forecast performance according to the $MCS$ criterion than the $FI(d)$ model. Yet, the $FI(d)$ models tend to lead in forecast performance as the horizon increases.


Finally, the figure allows us to compare the $HAR(3)$ model against the $AR(22)$ model. Note the crossing in preferred model according to the $MCS$ criterion between the $AR(22)$ and $HAR(3)$ models as both the forecast horizon, and degree of memory increase. The figure shows that for $h=5$, the $AR(22)$ model is always on top of the $HAR(3)$ model. Nonetheless, the preferred model between the two switches from the $AR(22)$ to the $HAR(3)$ model as the forecast horizon increases. Furthermore, the crossing happens sooner for higher degrees of memory. This suggests that the structure imposed by the $HAR(3)$ specification helps to improve forecasting performance for higher degrees of memory, and for larger forecast horizons, at the cost of lower performance at small horizons.

Overall, Table \ref{tab:DGP1}, and Figure \ref{plot:DGP1} extend the findings of previous studies on forecasting long memory when the long memory is generated by $ARFIMA$ processes. They show that high-order $AR$ models are good alternatives for short forecast horizons, while extending the analysis to show that $ARFIMA$ models are better suited for medium and large forecast horizons. Moreover, we find that the constraints imposed by the $HAR$ model improve forecasting performance over the unconstrained same-order $AR$ model for higher degrees of memory, and longer forecast horizons.

\subsection{{\textit{DGP}} 2: Cross-Sectional Aggregation}

\begin{figure}[h!]\caption{Proportion of times the top performing models are in the $MCS$ at a 95\% confidence level when forecasting $DGP$ 2 with different degrees of memory at several horizons.}\label{plot:DGP2}
	\hspace*{-1cm}   
	\includegraphics[scale=0.45]{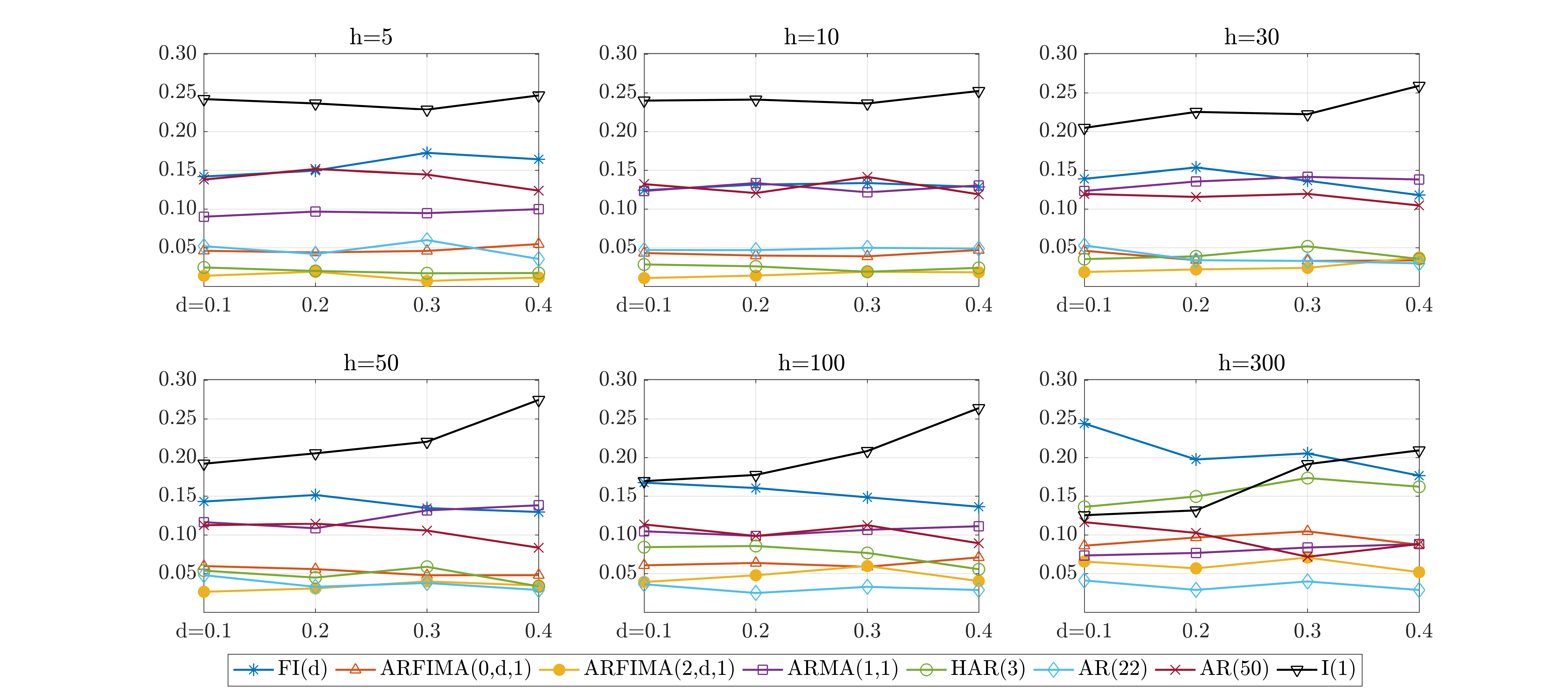}
\end{figure}

Results from the Monte Carlo analysis for the cross-sectional aggregated processes, $DGP$ 2, are presented in Table \ref{tab:DGP2}, and Figure \ref{plot:DGP2}. 

\begin{table}[h!]
	\centering
	\begin{scriptsize}
		\setlength{\tabcolsep}{3pt}
		\renewcommand{\arraystretch}{1} 
		\begin{tabular}{l|cc|cc|cc|cc|cc|cc}
			\hline
			$DGP$ 2	&\multicolumn{2}{c|}{h=5}	&\multicolumn{2}{c|}{10}		&\multicolumn{2}{c|}{30}		&\multicolumn{2}{c|}{50}	&\multicolumn{2}{c|}{100}	&\multicolumn{2}{c}{300}	\\
			$d=0.3$	&	$RMAD$	&	$MCS$	&	$RMAD$	&	$MCS$	&	$RMAD$	&	$MCS$	&	$RMAD$	&	$MCS$ &	$RMAD$	&	$MCS$ &	$RMAD$	&	$MCS$	\\
			\hline																	
			$FI(d)$	&	1.027	&	0.172	&	1.088	&	0.134	&	1.161	&	0.136	&	1.192	&	0.135	&	1.228	&	0.149	&	1.267	&\textbf{0.206}	\\
			$ARFIMA(1,d,0)$&\textbf{1.019}&	0.036	&\textbf{1.084}	&	0.034	&\textbf{1.159}	&	0.037	&\textbf{1.191}	&	0.037	&\textbf{1.227}	&	0.038	&\textbf{1.266}	&	0.062	\\
			$ARFIMA(0,d,1)$&	1.020	&	0.046	&	1.085	&	0.039	&\textbf{1.159}	&	0.032	&\textbf{1.191}	&	0.047	&\textbf{1.227}	&	0.059	&\textbf{1.266}	&	0.105	\\
			$ARFIMA(1,d,1)$&\textbf{1.019}&	0.020	&\textbf{1.084}	&	0.028	&	1.161	&	0.048	&	1.192	&	0.045	&\textbf{1.227}	&	0.041	&\textbf{1.266}	&	0.072	\\
			$ARFIMA(2,d,1)$&\textbf{1.019}&	0.007	&	1.086	&	0.019	&	1.164	&	0.024	&	1.196	&	0.040	&	1.230	&	0.060	&	1.268	&	0.071	\\
			\hline
			$ARMA(1,1)$	&	1.029	&	0.095	&	1.097	&	0.121	&	1.184	&	0.142	&	1.216	&	0.132	&	1.243	&	0.107	&	1.275	&	0.084	\\
			$ARMA(2,1)$	&	1.022	&	0.032	&	1.089	&	0.034	&	1.172	&	0.026	&	1.205	&	0.016	&	1.235	&	0.028	&	1.272	&	0.049	\\
			$ARMA(1,2)$	&	1.026	&	0.024	&	1.093	&	0.026	&	1.178	&	0.036	&	1.212	&	0.035	&	1.241	&	0.033	&	1.274	&	0.037	\\
			$ARMA(3,3)$	&	1.026	&	0.024	&	1.092	&	0.028	&	1.173	&	0.022	&	1.205	&	0.026	&	1.237	&	0.030	&	1.275	&	0.054	\\
			$ARMA(4,4)$	&	1.024	&	0.025	&	1.090	&	0.028	&	1.171	&	0.023	&	1.204	&	0.023	&	1.235	&	0.029	&	1.273	&	0.060	\\
			\hline
			$HAR(3)$	&	1.021	&	0.017	&	1.087	&	0.019	&	1.168	&	0.052	&	1.203	&	0.059	&	1.236	&	0.077	&	1.275	&	0.173	\\
			$AR(22)$	&	1.023	&	0.060	&	1.089	&	0.050	&	1.168	&	0.033	&	1.201	&	0.038	&	1.233	&	0.033	&	1.271	&	0.040	\\
			$AR(30)$	&	1.025	&	0.070	&	1.092	&	0.063	&	1.171	&	0.047	&	1.203	&	0.042	&	1.234	&	0.037	&	1.271	&	0.035	\\
			$AR(50)$	&	1.032	&	0.145	&	1.097	&	0.142	&	1.177	&	0.120	&	1.208	&	0.105	&	1.237	&	0.112	&	1.273	&	0.071	\\
			$I(1)$	&	1.075	&\textbf{0.227}	&	1.164	&\textbf{0.235}	&	1.275	&\textbf{0.222}	&	1.326	&\textbf{0.220}	&	1.387	&\textbf{0.208}	&	1.479	&	0.191	\\
			\hline																
		\end{tabular}
		\caption{Mean of the $RMAD$ and proportion of times the model is in the $MCS$ using the $MAD$ loss function and the $T_R$ statistic at a 95\% confidence level.}
		\label{tab:DGP2}
	\end{scriptsize}
\end{table}

Note that the $I(1)$ model is the one contained in the superior set of models the most for forecasts horizons up to 100, while performing last according to the $RMAD$ criterion. Once again, this seems to suggest a higher success rate for the $I(1)$ model at short and medium forecast horizons, but at the price of high variability on its performance, which may relate to the distance from the last observed value to the mean.

Looking for a better balance between the criteria, note that the $ARFIMA$ class of models is the preferred one according to the $RMAD$ criterion for all forecast horizons, with the $FI(d)$ in particular remaining among the top performing according to the $MCS$ criterion. In this sense, a weighted average between the criteria would presumably result in selecting the $FI(d)$ model as a well suited model for forecasting purposes, specially for medium and large forecast horizons. Among the short memory models, the $ARMA(1,1)$ and $AR(50)$ models could provide good forecast alternatives for short forecast horizons, as seen by their relatively high value for the $MCS$ criterion, and $RMAD$ relatively close to the minimum.

We can see the effect that the degree of long memory has on the results in Figure \ref{plot:DGP2}. We plot the percentage of number of times the models are contained in the $MCS$ for different degrees of memory, for all forecast horizons. 

The figure extends the findings in Table \ref{tab:DGP2}, it shows the good performance of the $ARMA(1,1)$ and $AR(50)$ models at short and medium forecast horizons, providing similar results to the $FI(d)$ specification. Nonetheless, the figure shows the increase in relative forecast performance of the $FI(d)$ model when the forecast horizon increases. Also, the plot shows the increase in forecast performance of the $HAR(3)$ model for large forecast horizons. In particular, while the performance in small forecast horizons is inferior in comparison to the unconstrained $AR(22)$, the constrains seem to introduce the additional structure needed for good medium and large horizon forecasts. Section \ref{Sec:Disc} will analyse this feature further. 

Overall, Table \ref{tab:DGP2} and Figure \ref{plot:DGP2} indicate that the $ARFIMA$ class of models are a good specification for forecast construction when working with long memory series generated by cross-sectional aggregation, $DGP$ 2, particularly for medium and large forecast horizons. This in the sense that they obtain a good balance performance among both evaluation criteria. Among the short memory models, the $AR(50)$ and $ARMA(1,1)$ could provide sensible alternatives for smaller forecasting periods. Finally, the $HAR(3)$ model starts to show good performance at larger horizons, with slightly inferior performance at short forecast horizons.

\subsection{{\textit{DGP}} 3: Error Duration Model}

Table \ref{tab:DGP4} presents the results from the Monte Carlo analysis for processes generated using the error duration model, $DGP$ 3, for long memory parameter $d=0.3$. 

\begin{table}[h!]
	\centering
	\begin{scriptsize}
		\setlength{\tabcolsep}{3pt}
		\renewcommand{\arraystretch}{1} 
		\begin{tabular}{l|cc|cc|cc|cc|cc|cc}
			\hline
			$DGP$ 3	&\multicolumn{2}{c|}{h=5}	&\multicolumn{2}{c|}{10}		&\multicolumn{2}{c|}{30}		&\multicolumn{2}{c|}{50}	&\multicolumn{2}{c|}{100}	&\multicolumn{2}{c}{300}	\\
			$d=0.3$	&	$RMAD$	&	$MCS$	&	$RMAD$	&	$MCS$	&	$RMAD$	&	$MCS$	&	$RMAD$	&	$MCS$ &	$RMAD$	&	$MCS$ &	$RMAD$	&	$MCS$	\\
			\hline																	
			$FI(d)$		&	1.097	&	0.130	&	1.123	&	0.126	&	1.131	&\textbf{0.162}	&	1.133	&\textbf{0.193}	&	1.136	&\textbf{0.217}	&	1.140	&\textbf{0.277}	\\
			$ARFIMA(1,d,0)$&	1.075	&	0.038	&	1.105	&	0.037	&	1.120	&	0.041	&	1.124	&	0.035	&	1.130	&	0.042	&	1.136	&	0.063	\\
			$ARFIMA(0,d,1)$&	1.074	&	0.028	&	1.109	&	0.028	&	1.130	&	0.027	&	1.137	&	0.023	&	1.145	&	0.029	&	1.156	&	0.059	\\
			$ARFIMA(1,d,1)$&\textbf{1.067}&	0.012	&\textbf{1.100}	&	0.012	&\textbf{1.118}	&	0.009	&\textbf{1.123}	&	0.021	&\textbf{1.129}	&	0.021	&\textbf{1.135}	&	0.037	\\
			$ARFIMA(2,d,1)$&\textbf{1.067}&	0.012	&\textbf{1.100}	&	0.016	&\textbf{1.118}	&	0.025	&\textbf{1.123}	&	0.021	&\textbf{1.129}	&	0.031	&\textbf{1.135}	&	0.041	\\
			\hline
			$ARMA(1,1)$	&	1.077	&	0.129	&	1.109	&	0.110	&	1.125	&	0.089	&	1.129	&	0.082	&	1.133	&	0.073	&	1.137	&	0.071	\\
			$ARMA(2,1)$	&	1.074	&	0.031	&	1.108	&	0.040	&	1.125	&	0.043	&	1.128	&	0.042	&	1.133	&	0.043	&	1.137	&	0.050	\\
			$ARMA(1,2)$	&	1.070	&	0.018	&	1.103	&	0.020	&	1.120	&	0.024	&	1.125	&	0.030	&	1.130	&	0.030	&	1.136	&	0.044	\\
			$ARMA(3,3)$	&	1.069	&	0.023	&	1.102	&	0.025	&	1.119	&	0.033	&	1.124	&	0.033	&	1.130	&	0.040	&	1.136	&	0.065	\\
			$ARMA(4,4)$	&	1.070	&	0.043	&	1.103	&	0.033	&	1.119	&	0.039	&	1.124	&	0.050	&	1.130	&	0.047	&\textbf{1.135}	&	0.079	\\
			\hline
			$HAR(3)$	&	1.076	&	0.100	&	1.106	&	0.104	&	1.121	&	0.086	&	1.126	&	0.095	&	1.131	&	0.099	&	1.137	&	0.145	\\
			$AR(22)$	&	1.077	&	0.052	&	1.108	&	0.062	&	1.122	&	0.074	&	1.126	&	0.066	&	1.131	&	0.057	&	1.136	&	0.065	\\
			$AR(30)$	&	1.078	&	0.074	&	1.109	&	0.077	&	1.124	&	0.094	&	1.126	&	0.089	&	1.131	&	0.085	&	1.136	&	0.096	\\
			$AR(50)$	&	1.085	&	0.146	&	1.116	&	0.149	&	1.131	&	0.138	&	1.133	&	0.125	&	1.134	&	0.134	&	1.137	&	0.143	\\
			$I(1)$	&	1.212	&\textbf{0.164}	&	1.261	&\textbf{0.161}	&	1.298	&	0.116	&	1.309	&	0.095	&	1.324	&	0.089	&	1.371	&	0.068	\\
			\hline																
		\end{tabular}
		\caption{Mean of the $RMAD$ and proportion of times the model is in the $MCS$ using the $MAD$ loss function and the $T_R$ statistic at a 95\% confidence level.}
		\label{tab:DGP4}
	\end{scriptsize}
\end{table}

We can see from the table that the $ARFIMA$ class of models provide the best performance measured by the $RMAD$ criterion for all forecast horizons, while showing the best performance by the $MCS$ criterion for $h=30$ and larger. Thus, the $ARFIMA$ class of models seem to provide the best forecast performance for series generated by the error duration model, $DGP$ 3. 

Turning to short memory alternatives, note the relatively good performance of the $AR(50)$ model for all forecast horizons. Thus, even though the $FI(d)$ model appears as the best performing model, high-order $AR$ models seem to produce good short and medium horizon forecast alternatives for $DGP$ 3. Moreover, contrasting the performance of the $HAR(3)$ model against the $AR(22)$ model, the table shows the gains in performance of imposing some structure into the higher-order $AR$ models when forecasting $DGP$ 3, with similar or better performance on both criteria. 

\begin{figure}[h!]\caption{Proportion of times the top performing models are in the $MCS$ at a 95\% confidence level when forecasting $DGP$ 3 with different degrees of memory at several horizons.}\label{plot:DGP4}
	\hspace*{-1cm}   
	\includegraphics[scale=0.45]{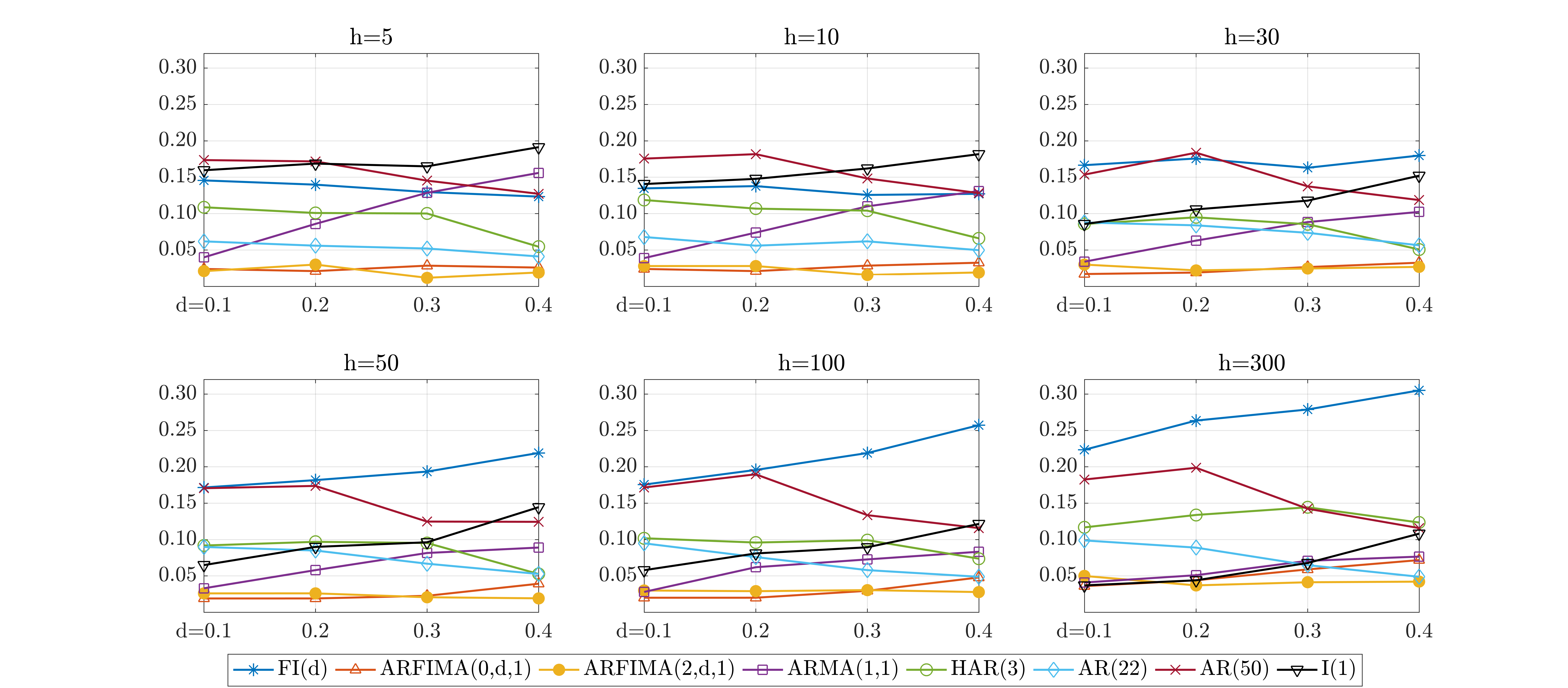}
	\vspace*{-0.4cm}
\end{figure}

Figure \ref{plot:DGP4} presents the proportion of times the models are contained in the $MCS$ when forecasting $DGP$ 3 for different degrees of memory. The figure shows the relative performance increase of the $FI(d)$ model over high-order $AR$ models as both the degree of memory, and the forecast horizon increase; thus, it extends the insights gained from the table to all degrees of memory considered. Also, note that the $HAR(3)$ model is always on top of the unconstrained $AR(22)$ model, if slightly for medium horizons.

Overall, Table \ref{tab:DGP4} and Figure \ref{plot:DGP4} convey one of the main insights from this paper; that is, short memory models are good alternatives for short horizon forecasts, while long memory models are preferred for medium and long horizon forecasts of long memory processes, regardless of the memory generating mechanism. 

\section{Discussion}\label{Sec:Disc}

Looking at the relative performance between an unconstrained high-order $AR$ model and a constrained $HAR$, the results from the Monte Carlo simulation can be further analysed in the context of the bias-variance trade-off typically studied in regression analysis.

All processes considered in this paper are long memory in the covariance sense; hence, the models are fitted to capture the information contained in the autocorrelation function and use it for forecasting purposes. In other words, the models select $\{a_i\}_{i=0}^T$ in the $x_t = a_0 + \sum_{i=1}^{T}{a_i x_{t-i}}$ representation, with the aim of replicating the autocorrelation function. 

$ARFIMA$ and $ARMA$ models differ in terms of the way to select the coefficients $a_{i}$. $ARFIMA$ models impose a hyperbolic rate by the fractional differencing operator $(1-L)^d$, see Equation \ref{eq:arfima}, while high-order $AR$ models are more flexible by selecting each coefficient individually. In this sense, the fractional models need just one parameter to establish the infinite list of coefficients, and are hence of low variance. Nonetheless, the uncertainty surrounding the estimation of the long memory parameter may introduce some bias. As an alternative, high-order $AR$ models are more flexible. Hence, they can reduce the bias of the coefficients, but suffer from increased variance given the number of estimated parameters. This distinction can be particularly important in the scenario of having small estimation samples, something we abstract from in this study. Given the uncertainty associated to estimating an increasing number of parameters, we would expect the performance of high-order $AR$ models to deteriorate in short series. Yet, as the Monte Carlo analysis showed, this flexibility can produce good forecast performance at short horizons, particularly when the degree of memory is small. Nonetheless, $AR$ models lose forecasting power as the forecast horizon gets larger. We could increase the order of the autoregressive process to incresae the forecasting performance at long horizons, but the estimation becomes unstable. 

In this context, $HAR$ models are a compromise between the rigid $ARFIMA$ and flexible high-order $AR$ model specifications. They incorporate high-order autoregressive specifications while greatly restricting the number of parameters to be estimated. This arrangement allows the $HAR$ model to provide similar forecast performance at medium forecast horizons as same-order unrestricted $AR$ models, while providing better long horizon forecasts. Yet, as shown in the Monte Carlo analysis, $HAR$ models may suffer a forecast performance loss at short horizons. 

\begin{figure}[ht!]\caption{Proportion of times the models are in the $MCS$ at a 95\% confidence level when forecasting $DGP$ 1 with different degrees of memory at several horizons. For the plots, the starting set contains only the six models shown.}\label{plot:HAR4}
	\hspace*{-1cm}   
	\includegraphics[scale=0.45]{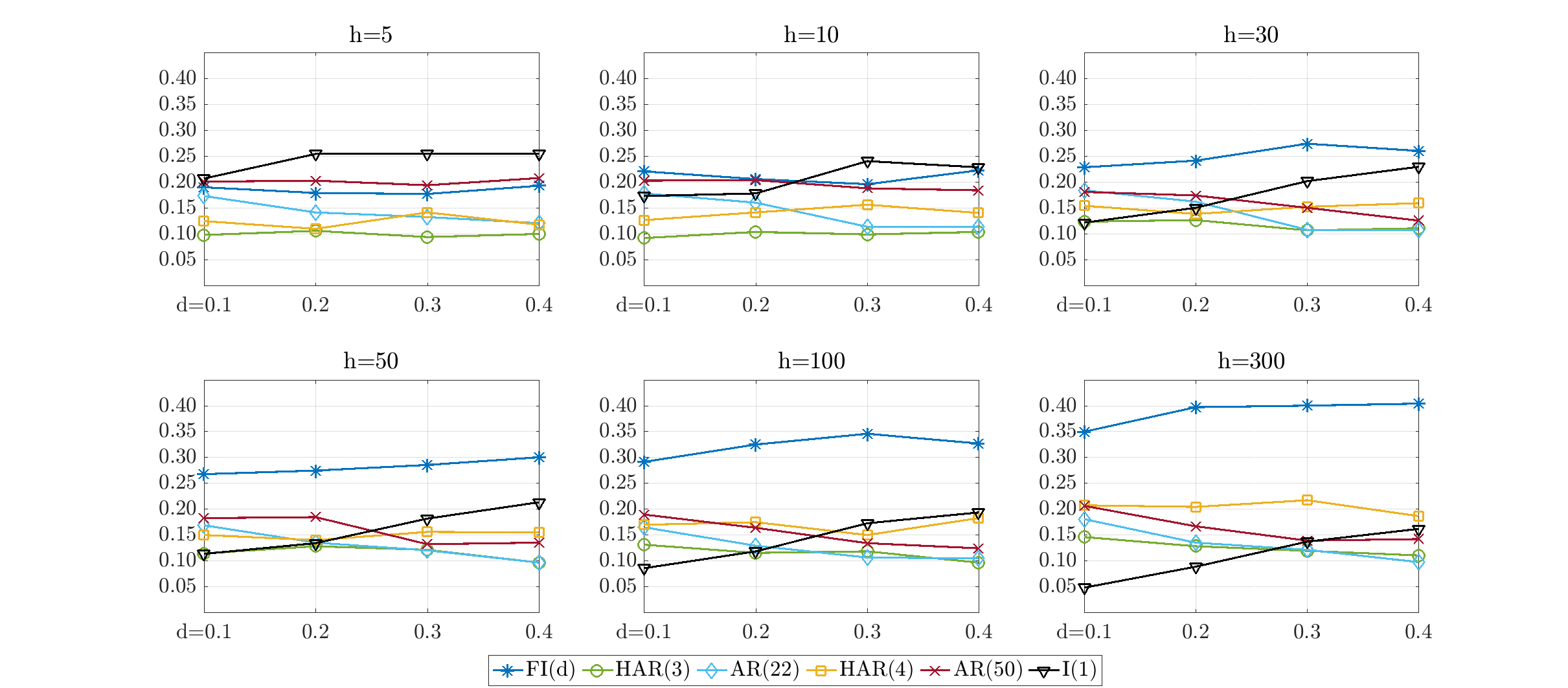}  
\end{figure}

To further illustrate this point, Figure \ref{plot:HAR4} compares the forecasting performance of constrained $AR$ models in the spirit of the $HAR$ model against their unconstrained specifications. In particular, in addition to the $HAR(3)$ and $AR(22)$ models, we show an unrestricted $AR(50)$ model and a $HAR(4)$ given by 
$$x_t = a_0+a_1x_{t-1}^{(f)}+a_2 x_{t-1}^{(w)} + a_3 x_{t-1}^{(m)}+a_4 x_{t-1}^{(b)}+\epsilon_t,$$ where $x_{t-1}^{(f)}=x_{t-1}$, $x_{t-1}^{(w)}=\frac{1}{5}\sum_{i=1}^5{x_{t-i}}$, $x_{t-1}^{(m)}=\frac{1}{22}\sum_{i=1}^{22}{x_{t-i}}$, and $x_{t-1}^{(b)}=\frac{1}{50}\sum_{i=1}^{50}{x_{t-i}}$. Note that the $HAR(4)$ model just described is a constrained $AR(50)$. 

\vspace*{0.2cm}

Figure \ref{plot:HAR4} shows the average number of times two high-order $AR$ processes and their comparable $HAR$ specifications are contained in the $MCS$ when forecasting $DGP$ 1. The figure displays the superior performance of unconstrained autoregressive processes against constrained ones of the same order for short forecast horizons. Nonetheless, it shows the increase in relative performance for the constrained versions at large forecast horizons. In particular, for $h=5$ and $h=10$, the unconstrained $AR$ models give better performance than equivalent order $HAR$ alternatives for all degrees of memory; while at $h=300$, the $HAR4$ is always on top of the unconstrained $AR(50)$.\footnote{The Online Appendix shows that this result extends to the other $DGP$s considered.} 

The bias-variance trade-off has been a topic of great interest in the literature of regressions with a high number of covariates, it thus would be compelling to adapt shrinkage and sparse methods to lag selection in the context of long memory forecasting. This line of inquiry is left open for future research.

\section{Illustrative Example}\label{Sec:App}

In this section, we evaluate the forecasting performance of the competing models on real data. We select the Realized Variance $(RV)$ of the S\&P 500 for illustration. As it is well known, the Realized Variance is a measure of volatility. We obtain the $RV$ series from the Oxford-Man Institute's ``Realised Library'' computed on the basis of intradaily observations spaced into 5-minute intervals and subsampled at a 1-minute frequency. The sample runs from January 3, 2000 until December 30, 2015.

The $RV$ has been proven to have long memory by, among others, \cite{Martens2009} and \cite{Andersen2003}. In particular, notice that by construction the S\&P 500 series is an aggregated measure; thus, it is in line with the cross-sectional argument for long memory. Furthermore, \cite{Parke1999} argues that given the difference between information quality among agents, the error duration model is capable of explaining the long memory in volatility. Hence, realized variance can be argued to have long memory by the theoretical explanations considered in this work, making it a good fit for the exercise. 

\begin{figure}[h!]\caption{Realized Variance and its autocorrelation function.}\label{plot:RV}
	\hspace*{1cm}   
	\includegraphics[scale=0.85]{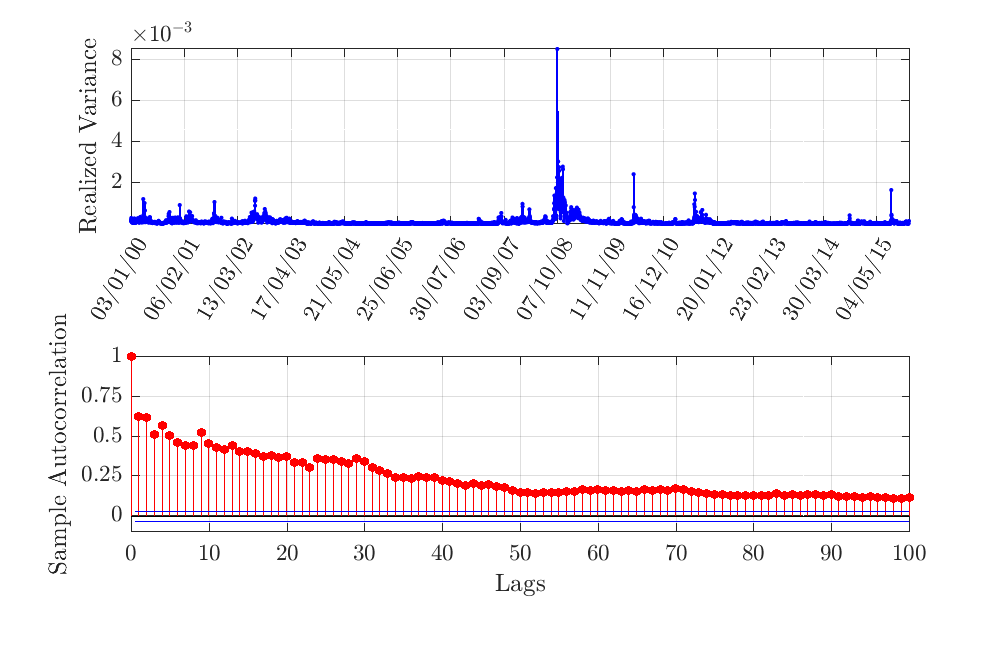}  
	\vspace*{-0.5cm}
\end{figure}

Figure \ref{plot:RV} presents the $RV$ series and its autocorrelation function. The autocorrelation function shows behaviour similar to those of long memory processes, remaining significant at large lags. The estimates for the long memory parameter are 0.4501, 0.4675, and 0.3756 by the semi-parametric estimator of \cite{Geweke1983}, the local Whittle estimator of \cite{Robinson1995} and \cite{Kunsch1986}, and the Maximum Likelihood Estimator of \cite{Sowell1992}, respectively. 

We use as estimation sample an increasing window starting from the period between January 3, 2000 to December 30, 2014. The last estimation window runs from January 3, 2000 to December 30, 2015. For each estimation window we construct forecasts for $5,10,22,66,120$ and 254 periods ahead, and compute the Model Confidence Set for each estimation window and forecast horizon.

Table \ref{tab:App} presents the results from the forecasting exercise on the first estimation window. As the table shows,\footnote{The $ARFIMA(2,d,1)$ showed convergence problems and thus was excluded from this exercise.} short memory models give the best forecast performance for horizons up to 22 periods ahead, in line with the results of \cite{Martens2009}. In particular, for short forecast horizons either the $ARMA(1,2)$ or the $ARMA(1,2)$ is the one with minimum $RMAD$, and the only one contained in the $MCS$. Nonetheless, the $ARFIMA$ type of models tends to dominate in forecast performance as the forecast horizon increases, in line with our results from the Monte Carlo analysis. For horizons $h\in\{66,120,254\}$, representing 3, 6 and 12 months, the $ARFIMA(0,d,1)$ is the only one contained in the $MCS$ and with lowest $RMAD$.

\begin{table}[h!]
	\centering
	\begin{scriptsize}
		\setlength{\tabcolsep}{3pt}
		\renewcommand{\arraystretch}{1} 
		\begin{tabular}{l|cc|cc|cc|cc|cc|cc}
			\hline
			Realized Variance	&\multicolumn{2}{c|}{h=5}	&\multicolumn{2}{c|}{10}		&\multicolumn{2}{c|}{22}		&\multicolumn{2}{c|}{66}	&\multicolumn{2}{c|}{120}	&\multicolumn{2}{c}{254}	\\
			&	$RMAD$	&	$MCS$	&	$RMAD$	&	$MCS$	&	$RMAD$	&	$MCS$	&	$RMAD$	&	$MCS$ &	$RMAD$	&	$MCS$ &	$RMAD$	&	$MCS$	\\
			\hline																	
			$FI(d)$	&	0.0049	&	0	&	0.0056	&	0	&	0.0060	&	0	&	\textbf{0.0054}	&	0	&	0.0053	&	0	&	0.0063	&	0	\\
			$ARFIMA(1,d,0)$	&	0.0051	&	0	&	0.0057	&	0	&	0.0060	&	0	&	0.0056	&	0	&	0.0057	&	0	&	0.0065	&	0	\\
			$ARFIMA(0,d,1)$	&	0.0054	&	0	&	0.0060	&	0	&	0.0063	&	0	&	\textbf{0.0054}	&	\textbf{1}	&	\textbf{0.0052}	&	\textbf{1}	&	\textbf{0.0062}	&	\textbf{1}	\\
			$ARFIMA(1,d,1)$	&	0.0054	&	0	&	0.0060	&	0	&	0.0063	&	0	&	\textbf{0.0054}	&	0	&	0.0053	&	0	&	0.0063	&	0	\\
			\hline
			$ARMA(1,1)$	&	0.0057	&	0	&	0.0064	&	0	&	0.0065	&	0	&	0.0060	&	0	&	0.0066	&	0	&	0.0076	&	0	\\
			$ARMA(2,1)$	&	0.0043	&	0	&	0.0044	&	0	&	\textbf{0.0048}	&	\textbf{1}	&	0.0077	&	0	&	0.0083	&	0	&	0.0086	&	0	\\
			$ARMA(1,2)$	&	\textbf{0.0041}	&	\textbf{1}	&	\textbf{0.0043}	&	\textbf{1}	&	0.0050	&	0	&	0.0078	&	0	&	0.0084	&	0	&	0.0086	&	0	\\
			$ARMA(3,3)$	&	0.0064	&	0	&	0.0065	&	0	&	0.0060	&	0	&	0.0074	&	0	&	0.0081	&	0	&	0.0085	&	0	\\
			$ARMA(4,4)$	&	0.0056	&	0	&	0.0058	&	0	&	0.0053	&	0	&	0.0075	&	0	&	0.0082	&	0	&	0.0085	&	0	\\
			\hline
			$HAR(3)$	&	0.0055	&	0	&	0.0057	&	0	&	0.0056	&	0	&	0.0068	&	0	&	0.0077	&	0	&	0.0083	&	0	\\
			$AR(22)$	&	0.0052	&	0	&	0.0057	&	0	&	0.0057	&	0	&	0.0069	&	0	&	0.0077	&	0	&	0.0083	&	0	\\
			$AR(30)$	&	0.0057	&	0	&	0.0064	&	0	&	0.0063	&	0	&	0.0064	&	0	&	0.0071	&	0	&	0.0079	&	0	\\
			$AR(50)$	&	0.0056	&	0	&	0.0062	&	0	&	0.0059	&	0	&	0.0070	&	0	&	0.0079	&	0	&	0.0084	&	0	\\
			$I(1)$	&	0.0074	&	0	&	0.0080	&	0	&	0.0083	&	0	&	0.0061	&	0	&	0.0053	&	0	&	0.0070	&	0	\\
			\hline																
		\end{tabular}
		\caption{$RMAD$ and indicator function for the model being contained in the $MCS$ using the $MAD$ loss function and the $T_R$ statistic at a 95\% confidence level.}
		\label{tab:App}
	\end{scriptsize}
\end{table}

\begin{figure}[ht!]\caption{Realized Variance and forecasts.}\label{plot:RV_For}
	\includegraphics[scale=0.44]{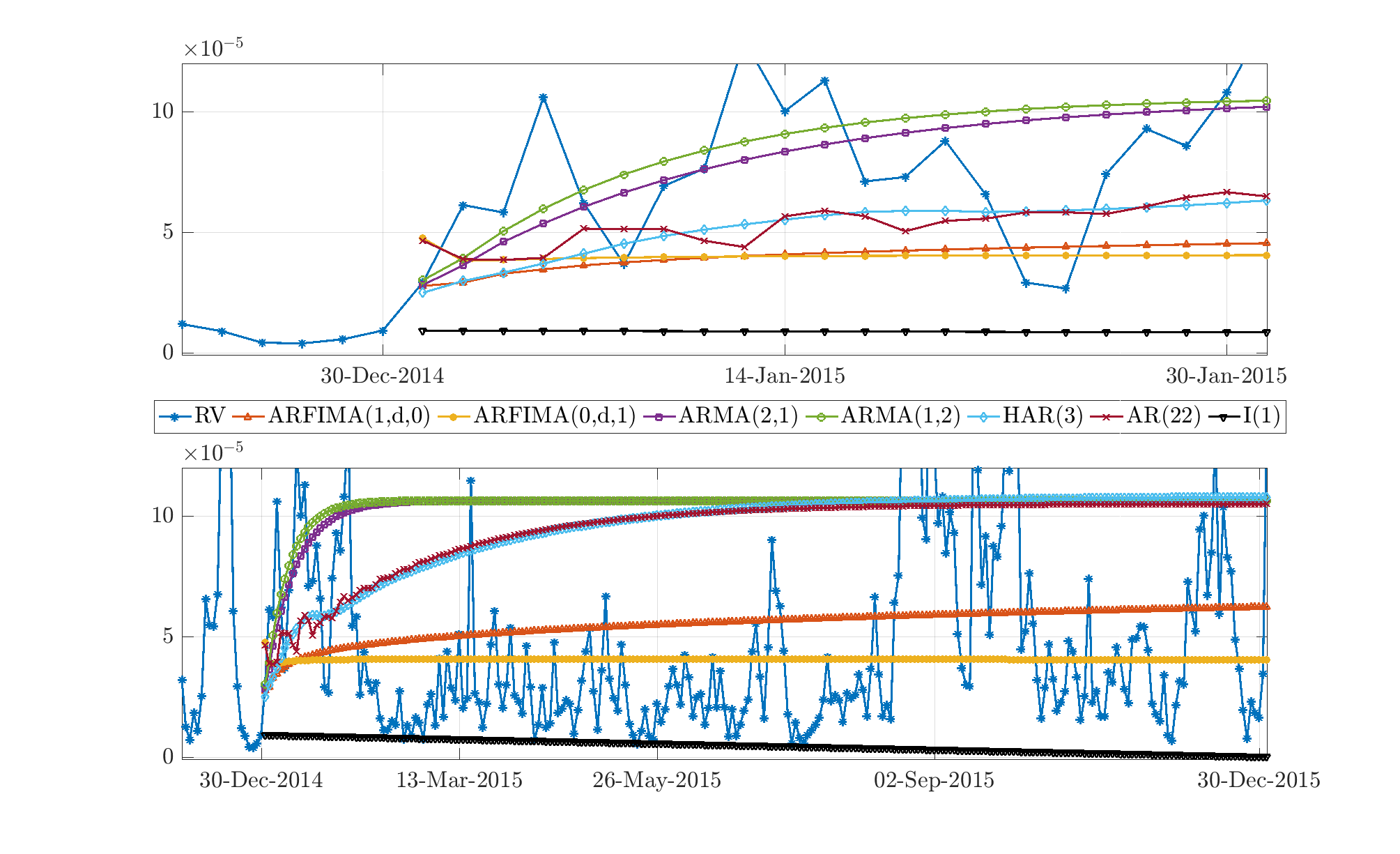}  
\end{figure}

The results from Table \ref{tab:App} can be better understood by looking at Figure \ref{plot:RV_For}. The figure shows the true $RV$ and the forecasts from the best competing models. The top plot shows the forecasts up to a month ahead, while the bottom one extends it to a year ahead. 

As the figure shows, all models determine that the $RV$ is away from its mean by the end of the estimation period. The main difference between models being the speed of convergence to the mean. The fast convergence of the $ARMA$ specifications allows them to capture the quick bounce at the beginning of the forecasting period much better than the $ARFIMA$ alternatives, which results in better short horizon forecasts. Nonetheless, the long memory alternatives are better suited to capture the long horizon dynamics of the $RV$ as the forecast horizon increases. 

For robustness, we extend the analysis to the exercise with an increasing estimation window. Table \ref{tab:App_Ext} presents the average number of times the competing models are in the superior set of models. For ease of exposition, we have pooled the results between long memory alternatives ($ARFIMA$ models) and short memory ones ($ARMA$ models).

\vspace*{0.3cm}

\begin{table}[ht!]
	\begin{small}
		\begin{center}
			\setlength{\tabcolsep}{3pt}
			\renewcommand{\arraystretch}{1} 
			\begin{tabular}{l|c|c|c|c|c|c}
				\hline
				Realized Variance	&$h=5$	&10		&22	&66	&120	&254\\
				\hline																	
				$ARFIMA$ alternatives & 0.268   & 0.342  &  \textbf{0.342} &   \textbf{0.425}  & \textbf{ 0.720}  &  \textbf{0.705} \\
				$ARMA$ alternatives &\textbf{0.409}    &\textbf{0.370}    &0.335   & 0.264  &  0.008       &  0 \\
				$I(1)$ no-change model &0.323    &0.287  &  0.323    &0.311   & 0.311  &  0.335 \\
				\hline						
			\end{tabular}
			\caption{Average number of times the type of model is in the $MCS$ at a 95\% confidence level for the increasing estimation window exercise. In total, 254 estimation windows were considered.}
			\label{tab:App_Ext}
		\end{center}
	\end{small}
\end{table}

The table confirms our simulations and first estimation window results. It shows that for the short horizon, the $ARMA$ models are the ones contained the most in the $MCS$. As we increase the forecasting horizon, the $ARFIMA$ alternatives are the ones contained the most in the $MCS$. In particular, no short memory alternative is contained in the $MCS$ for forecast up to a year ahead, for all estimation windows.

\section{Conclusions}\label{Sec:Con}

This paper argues that the most cited theoretical arguments behind the presence of long memory in the data do not correspond to the fractional difference operator. In this context, it evaluates the forecasting performance of $ARFIMA$ models when the memory is generated from nonfractional sources. 

We find that high-order $AR$ models produce comparable forecasts as $ARFIMA$ models at short horizons. Nonetheless, as the forecast horizon increases, the $ARFIMA$ models tend to dominate in terms of forecast performance. Hence, $ARFIMA$ models are well suited for medium and long horizon forecasts of long memory regardless of the generating mechanism, while high-order $AR$ models may be good alternatives for forecasts at short horizons. In particular, we find that if the long memory is generated by the error duration model, the fractionally integrated model produces the best forecast performance at medium and large horizons for all degrees of memory, while remaining competitive at short horizons.

Additionally, by making a compromise between flexibility and complexity, we find that the structure imposed by the $HAR$ model induces a trade-off in forecast performance at different forecast horizons. In other words, the $HAR$ model produces better long horizon forecasts, similar medium horizon forecasts, and similar or inferior short horizon forecasts, than same-order $AR$ model specifications.

Our results have implications for Climate Econometrics and Financial Econometrics models dealing with forecasts at different horizons. As an illustrative example, we show for the Realized Variance of the S\&P 500 that while short memory models are well suited for forecasts up to a month ahead, the $ARFIMA$ class of models dominate in forecast performance for longer horizons. 

\section*{Acknowledgements}
The author would like to thank Niels Haldrup for all the insightful comments, the paper greatly improved because of him.
The paper was written in part while on a visiting stay at Erasmus University Rotterdam. The author would like to thank the Faculty and Administrative Staff at the Erasmus School of Economics for all their help; in particular to Michel van der Wel and Dick van Dijk for the interesting conversations. 
This work was supported by CREATES - Center for Research in Econometric Analysis of Time Series (DNRF78), funded by the Danish National Research Foundation.

\section*{Appendix}
\renewcommand{\thesubsection}{\Alph{subsection}}

\subsection{Parameters}\label{app:par}
\begin{table}[h!]
	\centering
	\begin{footnotesize}
		\begin{tabular}{l|c}
			\hline
			All	&\begin{tabular}{c}$\varepsilon_t\sim i.i.d.N(0,1)$ $\forall t$ \\$T=1,000$; $R=1,000$\end{tabular}\\ 
			\hline
			$DGP$ 1	&$\phi_1 = 0.2$\\
			\hline
			$DGP$ 2 	&$N=10,000$; $p=1.4$\\
			\hline
			$DGP$ 3	&$p_k = (\Gamma(k+d)\Gamma(2-d))/(\Gamma(k+2-d)\Gamma(d))$\\
			\hline
		\end{tabular}
		\caption{Parameter configuration for the Monte Carlo analysis}\label{tab:params}
	\end{footnotesize}
\end{table}

\vspace*{1.5cm}
\subsection{Lag Selection Exercise}\label{app:lags}
\begin{table}[h!]
	\centering
	\begin{footnotesize}
		\begin{tabular}{l|c|c|c}
			\hline
			Model &$d$ &$ARFIMA$ &$ARMA$\\
			
			\hline
			$DGP$ 2	&0.2				&(1,0)	 	&(2,1)\\
			\cline{2-4}
			&0.4 		&(1,0)		&(2,1)\\
			\hline
			$DGP$ 3	&0.2		&(0,1)	&(1,1)	\\
			\cline{2-4}
			&0.4		&(0,1)	&(1,2)\\
			\hline
		\end{tabular}
		\caption{Results from the lag selection for the $AR$ and $MA$ components. We show the preferred model for each criteria from $R=1,000$ replications using a sample size of $T=1,000$. Results based on the Bayesian Information Criteria following the work of \cite{Beran1998} on consistency of information criteria for long memory processes.}\label{tab:lags}
	\end{footnotesize}
\end{table}


\end{document}